\documentclass[aps,amssymb,amsmath, pra]{revtex4}
\usepackage{graphicx}
\usepackage{epsfig}
\usepackage{dcolumn}
\usepackage{bm}
\usepackage{bbm}
\usepackage{color}
\usepackage{hyperref}
\usepackage[up]{subfigure}
\newcommand{\be}{\begin{equation}}
\newcommand{\ee}{\end{equation}}
\newcommand{\bc}{\begin{center}}
\newcommand{\ec}{\end{center}}
\newcommand{\bea}{\begin{eqnarray}}
\newcommand{\eea}{\end{eqnarray}}

\newcommand{\ra}{\rangle}

\begin{document}
\title{Spatial entanglement using a quantum walk on a many-body system}
\author{Sandeep K. \surname{Goyal}}
\email{goyal@imsc.res.in}
\affiliation{The Institute of Mathematical Sciences, CIT campus, Chennai 600 113, India}
\author{C. M. \surname{Chandrashekar}}
\email{cmadaiah@iqc.ca}
\affiliation{Institute for Quantum Computing, University of Waterloo, 
Ontario N2L 3G1, Canada}
\affiliation{Perimeter Institute for Theoretical Physics, Waterloo, ON, N2L 2Y5, Canada}
\begin{abstract}
The evolution of a many-particle system on a one-dimensional lattice, subjected
to a quantum walk can cause spatial entanglement in the lattice position,
which can be exploited for quantum information/communication purposes. We 
demonstrate the evolution of spatial entanglement and its dependence
on the quantum coin operation parameters, the number of particles present
in the lattice and the number of steps of the quantum walk on the
system. Thus, spatial entanglement can be controlled and optimized
using a many-particle discrete-time quantum walk.  
\end{abstract}
\maketitle
\preprint{Version}
\section{Introduction}
\label{intro}
Entanglement in a quantum state has been the fundamental resource in
many quantum information and computation protocols, such as
cryptography, communication, teleportation and algorithms \cite{NC00,
  HHH09}. To implement these protocols, generating an entangled state
is very important.  Similarly, studies on the interface between
condensed matter systems and quantum information have shown
entanglement as a signature of quantum phase transition \cite{ON02,
  OAF02, OROM06}. To understand the phases and  dynamics in many-body
systems an analysis of entanglement in many-body systems is very
important. 
Hence, various schemes have been proposed for entanglement generation
in quantum systems \cite{BH02, LHL03, RER07, WS09} and for understanding
entanglement in many-body systems \cite{AFOV08}.  
Quantum walk (QW) is one such process in which an uncorrelated
state can evolve to an entangled state and be used to analyze the
evolution of entanglement \cite{Kem03, CLX05}. 
\par
The QW, which was developed as a quantum analog of the classical random
walk (CRW), evolves a particle into an entanglement between its internal and position degrees of freedom. It has played a significant role in the development of quantum algorithms \cite{Amb03}.
 Furthermore, the QW has been used to demonstrate 
 coherent quantum control over atoms, quantum
 phase transition \cite{CL08}, to explain the phenomena such as
 breakdown of an electric field-driven system \cite{OKA05} and  direct
 experimental evidence for wavelike energy transfer within
 photosynthetic systems \cite{ECR07, MRL08}. Experimental
 implementation of the QW has also been  
reported \cite{DLX03, RLB05, PLP08, KFC09}, and various other schemes
have   
been  proposed  for  its  physical realization \cite{TM02, RKB02, EMB05, Cha06, MBD06}.
Therefore,  studying entanglement during the QW process will be useful
from a quantum information theory perspective and also contribute to
further investigation of the practical applications of the QW.  In this
direction, evolution of entanglement between single particle and
position with time (number of steps of the discrete-time QW) has been
reported \cite{CLX05}.  
\par
In this paper, we consider a multipartite quantum walk on a one-dimensional lattice and study the evolution of {\it spatial
  entanglement}, entanglement between different lattice points. All
the particles considered in the system are identical and
indistinguishable with two internal states (sides of the quantum
coin). Spatial entanglement generated using a QW can be controlled by
tuning different parameters, such as parameters in the quantum coin
operation, number of particles in the system and evolution time (number of
steps). To quantify entanglement in the system we are
using Meyer-Wallach multipartite entanglement measure. 
\par
In Sec. \ref{qw}, we describe single-particle and many-particle
discrete-time QWs. In Sec. \ref{entanglement}, entanglement between a particle and  
position space and spatial entanglement using single- and many-particle
QWs are discussed. In Sec. \ref{mpent}, we present the measure for
spatial entanglement of the system using the Meyer-Wallach global
entanglement measure scheme for particles in a one-dimensional lattice
and in a closed chain ($n-$cycle). We also demonstrate control
over spatial entanglement by
exploiting the dynamical properties of the QW.   
We conclude with the summary in Sec. \ref{conc}. 
\section{Quantum Walk}
\label{qw}
Classical random walk (CRW) describes the dynamics of a particle in position
space with a certain probability. The QW is the quantum analog of CRW-developed exploiting 
features  of quantum
mechanics such as superposition and interference of quantum amplitudes
\cite{GVR58, FH65, ADZ93}. The QW,  which involves superposition of
states, moves simultaneously exploring multiple possible paths with
the amplitudes corresponding to the different  paths interfering. This
makes the variance of the QW on a line to grow quadratically with the
number of steps which is in sharp contrast to the linear growth for
the CRW.   
\par
The study of QWs has been largely divided into two standard variants:
 discrete-time QW (DTQW) \cite{ADZ93, DM96, ABN01} and a continuous-time QW (CTQW)  \cite{FG98}.  
In the CTQW, the walk is defined directly on the {\it position} Hilbert
space  $\mathcal{H}_p$, whereas for the DTQW it is necessary to
introduce an additional {\it coin} Hilbert space $\mathcal{H}_c$, a quantum
coin operation to define the direction in which the particle amplitude
has to evolve. The connection between these two variants and the generic
version of the QW has been studied \cite{FS06,
  C08}. However, the coin degree of freedom in 
the DTQW is an advantage over the CTQW as it allows control of dynamics of the QW \cite{AKR05, CSL08}. Therefore, we take full advantage of the coin degree of freedom in this work and
study the DTQW on a many-particle system. 

\subsection{Single-particle quantum walk}
\label{spqw}
The DTQW is defined on the Hilbert space $\mathcal  H=  \mathcal H_{c}
\otimes \mathcal H_{p}$. In one dimension, the coin Hilbert space $\mathcal H_{c}$,
spanned  by the basis state $|0\rangle$ and $|1\rangle$, represents two
sides of the quantum coin, and the position Hilbert space $\mathcal H_{p}$,
spanned by the basis states $|\psi_j\rangle$, $j  \in \mathbb{Z}$,
represent the positions in the lattice. To implement the DTQW, we will
consider a three-parameter U(2) operator $C_{\xi, \theta, \zeta}$ of
the form 
\be 
\label{coin}
C_{\xi,\theta,\zeta}
\equiv    \left(   \begin{array}{clcr}   e^{i\xi}\cos(\theta)    &   &
e^{i\zeta}\sin(\theta)    \\     e^{-i\zeta}    \sin(\theta)    &    &
-e^{-i\xi}\cos(\theta)
\end{array} \right)
\ee
as the quantum coin operation \cite{CSL08}.  The quantum coin operation is applied on the particle state ($C_{\xi,   \theta,  \zeta}  \otimes   {\mathbbm 1}$) when the initial state of the complete system is
\be
\label{qw:in}
|\Psi_{in}\rangle= \left [ \cos(\delta)|0\rangle +
  e^{i\eta}\sin(\delta)|1\rangle \right ] \otimes |\psi_{0}\rangle. 
\ee
The state $\cos(\delta)|0\rangle +  e^{i\eta}\sin(\delta)|1\rangle$ is the
state of the particle and $|\psi_{0}\rangle$ is the state of the
position at the lattice position $j=0$. 
\par
The quantum coin operation on the particle is followed by the conditional unitary shift operation $S$ which acts on the complete
Hilbert space of the system: 
\be 
\label{eq:alter} 
S =\exp(-i \sigma_{z}\otimes Pl), 
\ee
where $P$ is the  momentum operator, $\sigma_{z}$ is the Pauli
spin operator in the $z$ direction and $l$ is the length of each step.  The
eigenstates of $\sigma_{z}$ are denoted by $|0\rangle$ and
$|1\rangle$.  
Therefore, $S$, which delocalizes the wave packet over the positions
$(j-1)$ and $(j+1)$, can also  be written as 
\begin{eqnarray}
\label{eq:condshift}  S  =  |0\rangle  \langle 0|\otimes  \sum_{j  \in
\mathbb{Z}}|\psi_{j-1}\rangle  \langle \psi_{j} |+|1\rangle  \langle 1
|\otimes \sum_{j \in \mathbb{Z}} |\psi_{j+1}\rangle \langle \psi_{j}|.
\end{eqnarray}
\par
\begin{figure}[ht]
\begin{center}
\epsfig{figure=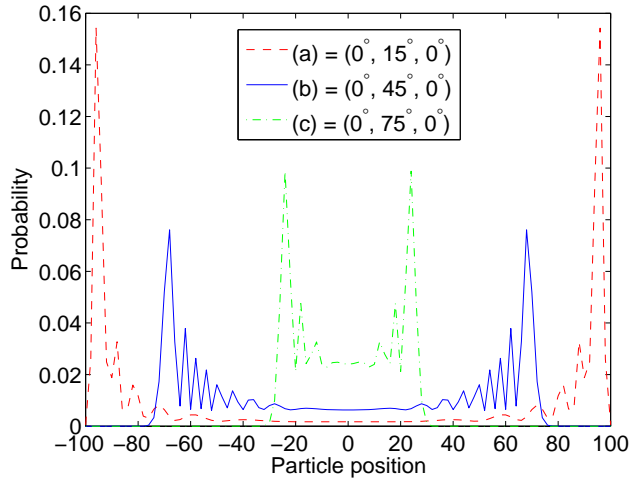, width=9.0cm}
\caption{\label{fig:qw1a}(color online) Spread of the probability distribution for
  different values of $\theta$ using the quantum coin operator $C_{0,
    \theta, 0}$. The distribution is wider for (a) $(0,
    \theta, 0)= (0, \frac{\pi}{12},  0)$ than for (b) $(0,
    \theta, 0)= (0,  \frac{\pi}{4}, 0)$ and (c)
  $(0, \theta, 0)= (0, \frac{5 \pi}{12}, 0)$, showing the decrease in spread with
  increase in $\theta$. The initial state of the particle is
  $|\Psi_{ins}\rangle = \frac{1}{\sqrt 2}\left ( |0\rangle + i
    |1\rangle \right ) \otimes |\psi_{0}\rangle$ and the distribution is
  for 100 steps.} 
\end{center}
\end{figure}

The process of
\be
\label{dtqwev}
 W_{\xi, \theta, \zeta} =
S(C_{\xi,   \theta,  \zeta}  \otimes   {\mathbbm 1})
\ee
is iterated without resorting to intermediate
measurement to help realize a large number of steps of the QW. The parameters  
$\delta$ and  $\eta$ in Eq. (\ref{qw:in}) can be  varied to obtain
different 
initial states of the particle. The three parameters $\xi$,
$\theta$  and $\zeta$ of $C_{\xi, \theta,
  \zeta}$ can be varied to  choose the quantum coin operation. By
varying parameter $\theta$ the variance can be 
increased  or  decreased according to the functional  form, $\sigma^{2}  \approx
(1-\sin(\theta))t^{2}$, where $t$ is the number of steps of the QW, as shown in Fig.  \ref{fig:qw1a}. 
\par
{\it Biased coin operation and biased QW:}  The most widely studied
form of the DTQW is the walk using the Hadamard operation  
\be
H = \frac{1}{\sqrt{2}}
  \begin{pmatrix}
    1 & \mbox{~} 1 \\
    1 & -1
  \end{pmatrix},
\label{hadamard}
\ee
corresponding to the quantum coin operation with $\xi = \zeta = 0$ and $\theta = \pi/4$
in Eq. (\ref{coin}). The Hadamard operation is an unbiased coin operation, and the resulting walk is  known as the Hadamard walk. This walk  implemented on a
particle initially in a symmetric superposition state,
\be
\label{qw:in1}
|\Psi_{ins}\rangle = \frac{1}{\sqrt 2} \left [ |0\rangle + i |1\rangle
\right ] \otimes |\psi_{0}\rangle, 
\ee 
 obtained by choosing $\delta = \pi/4$ and $\eta = \pi/2$ in
 Eq. (\ref{qw:in}), returns a symmetric, unbiased probability
 distribution of the particle in position space. However, the
 Hadamard walk on any asymmetric initial state of the particle results
 in an asymmetric, biased probability 
 distribution of the particle in position space
 \cite{ABN01}. We should note that the role of the initial state on the
 symmetry of the probability distribution is not vital for a QW using the 
 three-parameter operator given by Eq. (\ref{coin}) as a quantum coin
 operation.  
 \par
\begin{figure}[ht]
\begin{center}
\epsfig{figure=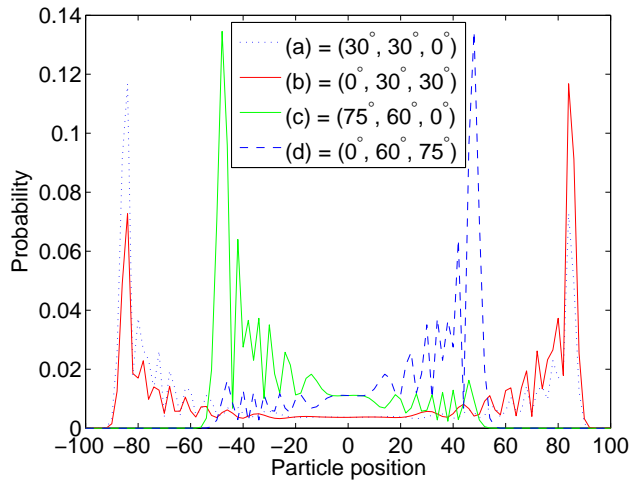, width=9.0cm}
\caption{Spread of probability distribution for different values of
  $\xi$, $\theta$, $\zeta$ using the quantum coin operator $C_{\xi,
    \theta, \zeta}$. The parameter $\xi$ shifts the distribution to the
  left: (a)$(\xi, \theta, \zeta) = ( \frac{\pi}{6}, \frac{\pi}{6}, 0)$
  and (c) $(\xi, \theta, \zeta)= (\frac{5 \pi}{12}, \frac{\pi}{3}, 0
  )$. The parameter $\zeta$  shifts it to the right: (b) $(\xi, \theta,
  \zeta)= (0, \frac{ \pi}{6}, \frac{\pi}{6})$ and  (d) $(\xi, \theta,
  \zeta) = (0, \frac{\pi}{3},  \frac{5 \pi}{12})$. The initial state
  of the particle $|\Psi_{ins}\rangle = \frac{1}{\sqrt{2}}(|0\rangle +
  i |1\rangle) \otimes |\psi_{0}\rangle$ and the distribution is for
  100 steps.} 
\label{fig:qw2}
\end{center}
\end{figure}

To elaborate this further, we will consider the first-step evolution
of the DTQW using a three-parameter quantum coin operation given by
Eq. (\ref{coin}) on a particle initially in the symmetric
superposition state.  
After the first step of the DTQW the state can be written as  
\begin{eqnarray}
\label{eq:condshift2}
W_{\xi, \theta, \zeta}|\Psi_{ins}\rangle =  \frac{1}{\sqrt 2}
\left [ \left(e^{i\xi}  \cos(\theta)+ i e^{i\zeta} \sin(\theta)\right )
|0\rangle|\psi_{-1}\rangle 
+ \left( e^{-i\zeta}\sin (\theta) - i e^{-i\xi} 
\cos(\theta)\right) |1\rangle|\psi_{+1}\rangle \right ]. 
\end{eqnarray}
If $\xi=\zeta$, Eq. (\ref{eq:condshift2}) has left-right symmetry in
the position probability distribution,  but not otherwise. That is,
the parameters $\xi$ and $\zeta$ introduce asymmetry in the position
space probability distribution.  Therefore, a coin operation with $\xi
\neq \zeta$ in Eq. (\ref{coin}) can be called as a biased quantum coin
operation which will bias the QW probability distribution of the particle 
initially in a symmetric superposition state (Fig. \ref{fig:qw2})
\cite{CSL08}.  However, we should note that irrespective of the
quantum coin operation used, QW can also be biased by 
choosing an 
asymmetric initial state of the particle (for example, the Hadamard walk
of a particle initially in the state $|0\rangle$ or the state $|1\rangle$).  
\subsection{Many-particle quantum walk}
\label{mbqw1}
\begin{figure}[ht]
\begin{center}
\includegraphics[width=8.5cm]{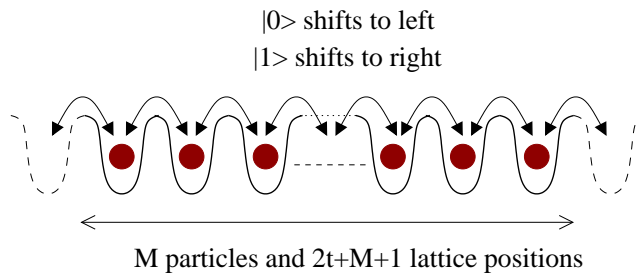}
\caption{Many-particle state with one non-interacting particle at each
  position space.} 
\label{mi}
\end{center}
\end{figure}

To define a many-particle QW in one dimension, we will consider an $M$-particle system  with one non-interacting particle at each position
(Fig.  \ref{mi}). The $M$ identical particles in $M$ lattice points with
each particle having its own coin and position Hilbert space will have
a  total Hilbert space $\mathcal{H}=\left( 
  \mathcal{H}_c \otimes \mathcal{H}_p \right)^M $. We
assume the particles to be distinguishable.  
\par
The evolution of each step of the QW on the $M$-particle system is given by
the application of the operator
$W_{0,\theta, 0}^{\otimes M}$. 
The initial state that we will consider for the many-particle system in one dimension will be    
\begin{equation}
  |\Psi_{ins}^{M}\rangle = \bigotimes_{j=-\frac{M-1}{2}}^{j=\frac{M-1}{2}}
  \left( \frac{|0\rangle + i|1\rangle}{\sqrt{2}} \right) \otimes
  |\psi_{j}\rangle. 
  \label{initialMBQWstate}
\end{equation}
\begin{figure}[ht]
\includegraphics[width=9.0cm]{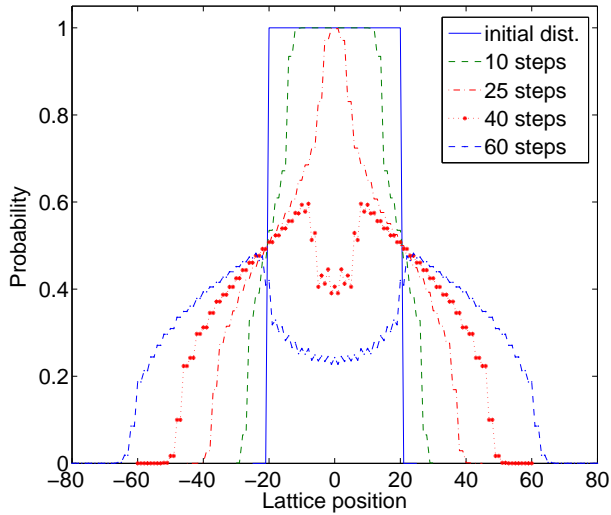}
\caption{(color online) Probability distribution of 40 particles
  initially with one particle in each position space when subjected to
  the QW of different number of steps. The initial state of all the
  particles is $\frac{1}{\sqrt 2}(|0\ra + i |1\ra)$ and is evolved in
  position space using the Hadamard operator, $C_{0, \pi/4, 0}$  as
  the quantum coin. The distribution spreads in the position space with
  an increase in number of steps.} 
\label{miqw}
\end{figure}
\par
For an $M$-particle system after $t$ steps of the QW, the
  Hilbert space consists of the tensor product of single lattice
  position Hilbert space which is $(2t+M+1)$ in number. That
  is, after $t$ steps of the QW, the $M$ particles are spread  
  between $(j-t)$ to $(j+t)$. In principle, each lattice point is associated
with a Hilbert space spanned by two subspaces, a zero-particle
subspace  and one-particle subspace spanned by two possible states of
the coin, $|0\rangle$ and $|1\rangle$.  
Therefore, the dimension of each lattice point will be $3^{M}$  and the
dimension of total Hilbert space is $(3^{M})^{\otimes M}$. 
Fig. (\ref{miqw}) shows the probability distribution of the many-particle 
system with an increase in number of steps of the QW.
\section{Entanglement}
\label{entanglement}
To efficiently make use of entanglement as a physical resource, the
amount of entanglement in a given system has to be
quantified. Therefore, entanglement in a pure bipartite system or a
system with two Hilbert spaces is quantified using standard measures
known as entropy of entanglement or  Schmidt number
\cite{NC00}. The entropy of entanglement corresponds to the von
Neumann entropy, a functional of the eigenvalues of the reduced density
matrix, and a Schmidt number is the number of non-zero Schmidt
coefficients in its Schmidt decomposition. For a multipartite state,
there are quite a few good entanglement measures that have been proposed \cite{CKW00,
  BL01,  EB01, MW02, VDM03, Miy08, HJ08}. However, as the number of
particles in the system increases, the complexity of finding an
appropriate entanglement measure also increases, making scalability
impractical.  Among the proposed measures, to address
this scalability 
problem, Mayer and Wallach proposed a {\em scalable} global entanglement
measure (polynomial measure) to quantify entanglement in many-particle
systems \cite{MW02}. 
\par 
 In this section, we will first discuss the entanglement of a 
 particle with position space quantified using entropy of
 entanglement. Later we will discuss spatial
 entanglement quantified using the Mayer-Wallach (M-W) measure. Spatial
 entanglement has been explored earlier using different methods. For
 example, in an ideal bosonic gas it has been studied using
 off-diagonal long-range order  \cite{HAKV07}. For our investigations,
 we consider a distinguishable many-particle system, implement QW and
 use the M-W measure to quantify spatial entanglement. In this system the
 dynamics of particles can be controlled by varying the quantum coin
 parameters, the initial state of the particles, the number of particles in
 the system and the number of steps of the QW. In particular, we choose
 the particles in one-dimensional open and closed chains. The spatial
 entanglement thus created can be used for example to create
 entanglement between distant atoms in an optical lattice \cite{SGC09}
 or as a channel for state transfer in spin chain systems \cite{Bos03,
   CDE04, CDD05}.   

\subsection{Single-particle - position entanglement}
\label{cpent}

QW entangles the particle (coin) and the position degrees of freedom. 
To quantify it, let us consider a DTQW on a particle
initially in a state given by 
Eq. (\ref{qw:in1}) with a simple form of a coin operation
\be 
\label{coin1}
C_{0,\theta, 0}
\equiv    \left(   \begin{array}{clcr}   \cos(\theta)    &   &
\sin(\theta)    \\        \sin(\theta)    &    &
-\cos(\theta)
\end{array} \right).
\ee
After the first step, $W_{0,\theta,0} = S(C_{0, \theta, 0} \otimes
{\mathbbm 1})$,  the state takes the form 
\begin{eqnarray}
  | \Psi_{1} \rangle &=& W_{0,\theta, 0} | \Psi_{ins}\rangle   
                   = \gamma \left( |0 \rangle \otimes |\psi_{j-1} \rangle
                   \right) + \delta \left( |1 \rangle \otimes |\psi_{j+1}
                     \rangle \right) 
  \label{evolution01}
\end{eqnarray}
where $\gamma = \left( \frac{\cos(\theta) + i\sin(\theta)}{\sqrt{2}} \right)$ and
$\delta = \left( \frac{\sin(\theta) - i\cos(\theta)}{\sqrt{2}} \right)$. 
The Schmidt rank of   $|\Psi_1\rangle$ is
$2$ which implies entanglement in the system. 
The value of entanglement with an increase in the number of steps can be further quantified  by  
computing the von Neumann  entropy of the reduced density matrix of
the position subspace.  
\begin{figure}[ht]
\includegraphics[width=9.0cm]{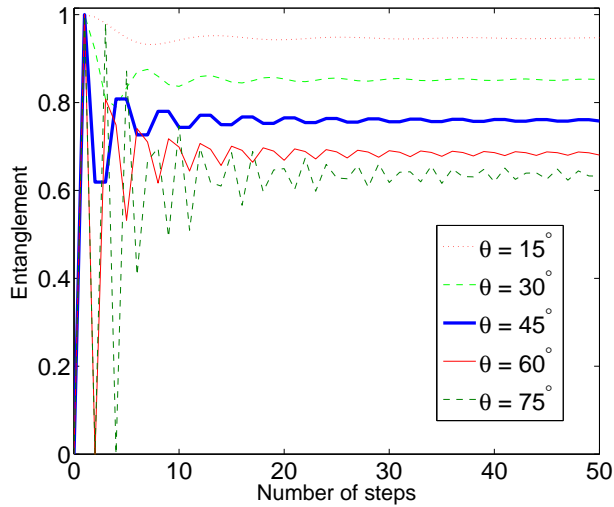}
\caption{(color online) Entanglement of a single particle with position
  space when subjected to the QW. The initial state of a particle is
  $\frac{1}{\sqrt 2}(|0\ra + i |1\ra)$ and is evolved in position
  space using different values for $\theta$ in the quantum coin
  operation $C_{0, \theta, 0}$. The entanglement initially oscillates
  and approaches an asymptotic value with an increase in the number of
  steps. For smaller values of $\theta$ the entanglement is higher and
  decreases with an increase in $\theta$. Initial oscillation is also
  larger for higher $\theta$.} 
\label{enta}
\end{figure}
\begin{figure}[ht]
\includegraphics[width=9.0cm]{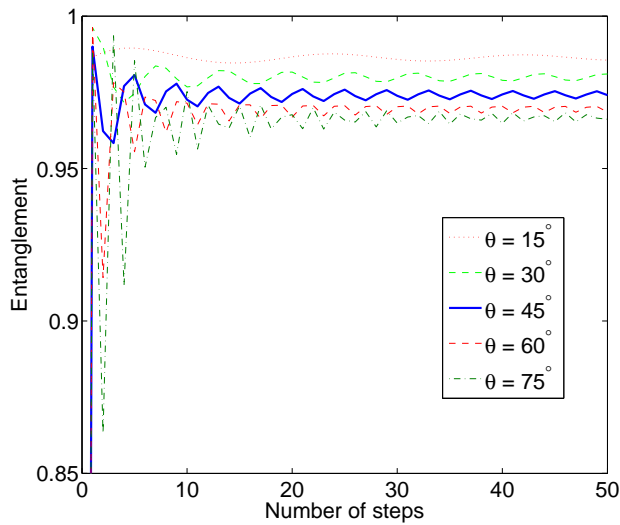}
\caption{(color online) Entanglement of single particle with position
  space when subjected to the QW. The initial state of the particle is
  given by Eq. (\ref{qw:in}) with $\delta = \frac{2\pi}{9}$ and $\eta
  = \frac{\pi}{6}$ and is evolved in position space using different
  values for $\theta$ in the quantum coin operation $C_{0, \theta,
    0}$. The entanglement initially oscillates and approaches an
  asymptotic value with an increase in the number of steps. For smaller
  values of $\theta$ the entanglement is higher and decreases with an 
  increase in $\theta$. Initial oscillation is also larger for higher
  $\theta$.} 
\label{enta1}
\end{figure}
\par
Fig. \ref{enta}  shows a plot of the entanglement against the number of
steps of the QW on a particle initially in a symmetric
superposition state using different values for $\theta$ in the
operation $W_{\theta}$. The von Neumann entropy of the reduced density
matrix of the coin is used to quantify entanglement between the
coin and the position in Fig. \ref{enta}. That is, 
\be
E_{c}(t) = - \sum_{j} \lambda_{j} \rm{log}_{2}(\lambda_{j})
\ee
where $\lambda_{j}$ are eigenvalues of the reduced density matrix of
the coin after $t$ steps (time).  The entanglement initially
oscillates and 
reaches an asymptotic value 
with increasing number of steps.  In the asymptotic limit,
the entanglement value decreases with an increase in $\theta$ and this
dependence can be attributed to the spread of the amplitude distribution
in position space.  That is, with an increase in $\theta$, constructive
interference of quantum amplitudes toward the origin becomes prominent
narrowing the distribution in the position space. 
In Fig.  \ref{enta1}, the process is repeated for a particle initially
in an asymmetric superposition state $|\Psi_{in}\rangle =
\left [\cos(\frac{2\pi}{9}) |0\rangle + e^{i \frac{\pi}{6}}
\sin(\frac{2\pi}{9})|1\rangle \right ] \otimes |\psi_{0}\rangle$.
Comparing Fig. \ref{enta1} with Fig. \ref{enta}, we can note the
increase in entanglement and decrease in the oscillation. This
observation can be explained by going back to our earlier note on
biased QW in Sec. \ref{spqw}. In Fig. \ref{fig:qw2} we note that
biasing of the coin operation leads to an asymmetry in the probability
distribution, with an increase in peak height on one side and a decrease
on the other side (increase and decrease are in reference to the symmetric
distribution). A similar biasing effect can also be reproduced by
choosing an asymmetric initial state of the particle. The biased
distribution with an increased value of probability at one side in the
distribution contributes to a reduced oscillation in the 
distribution. This  in turn results in the increase
of  the von Neumann entropy: entanglement.  
\subsection{Spatial entanglement}
\label{spentqw}
{\em Spatial entanglement} is the entanglement between the lattice
points. This entanglement takes the form of non-local particle number
correlations between spatial modes.  To observe spatial entanglement
we first need to associate the lattice 
with the state of a particle.  Then we need to consider the evolution of
a single-particle QW followed by the evolution of a many-particle QW, in order to
understand spatial entanglement. 
\subsubsection{Using a single-particle quantum walk}
\label{spentqw1}
In a single-particle QW, each lattice point is associated
with a Hilbert space spanned by two subspaces. The first is the
zero-particle subspace which does not involve any coin (particle)
states. The other is the one-particle subspace spanned by the two
possible states 
of the coin, $|0\rangle$ and $|1\rangle$.  To obtain the spatial
entanglement we will write the state of
  the particle in the form of the state of a lattice. Following from
  Eq. (\ref{evolution01}), the state of the 
  particles after 
first two steps of QW takes the form
  \begin{eqnarray}
    | \Psi_{2} \rangle &=& W_{0,\theta, 0} | \Psi_{1} \rangle  =  \gamma \left [
      \cos (\theta) |0\rangle |\psi_{j-2} \rangle  + \sin (\theta)
      |1\rangle|\psi_{j} \rangle 
    \right ]  + \delta \left [  \sin(\theta) |0\rangle |\psi_{j} \rangle -
  \cos(\theta) |1\rangle|\psi_{j+2} 
      \rangle \right ].
    \label{evolution02}
  \end{eqnarray}
  In order to obtain the state of the lattice we can redefine the position
 state in the  
following way: the occupied position state $|\psi_j\rangle$ as
$|1_{j}\rangle$, which means  
that the $j^{th}$ position is occupied and the rest of the lattice is
empty. Therefore, we can rewrite Eq. (\ref{evolution02}) as 
\begin{eqnarray}
    | \Psi_{2} \rangle  &  = & \gamma \left [
      \cos (\theta) |0\rangle | 1_{j-2} \rangle  + \sin (\theta)
      |1\rangle| 1_{j} \rangle 
    \right ]     + \delta \left [  \sin(\theta) |0\rangle | 1_{j} \rangle -
  \cos(\theta) |1\rangle| 1_{j+2} 
      \rangle \right ].
    \label{evolution03}
  \end{eqnarray}
Since we are interested in the spatial entanglement, we project this
state into one of the coin state so that we can ignore the
entanglement between the coin and the position state and consider only the
lattice states. 
Here we will choose the coin state to be $|0\rangle$ and take
projection to obtain the state of the lattice in the form 
 \begin{equation}
  |\Psi_{lat}\rangle  =  |0\rangle \left(\gamma
    \cos(\theta)|1_{j-2}\rangle + \delta \sin(\theta) 
    |1_{j}\rangle
  \right). 
  \label{latticestate}
\end{equation}
Each lattice site $j$ can be considered as a Hilbert space with basis states 
$|1_{j}\rangle$ (occupied state) and $|0_{j}\rangle$
(unoccupied state). Then, the above Eq. (\ref{latticestate}) in
the extended Hilbert space of each lattice can be rewritten in terms
of occupied and unoccupied lattice states as 
 \begin{equation}
  |\Psi_{lat}^{\prime}\rangle =  \gamma \cos(\theta) |1_{j-2}
  \,0_{j}\rangle+\delta \sin(\theta) |0_{j-2} \,1_{j}  \rangle.  
  \label{latticestate1}
\end{equation}
We can see that after first two steps of the QW the lattice points $j$ and
$(j-2)$ are entangled. One 
  can check that the lattice points $j$ and $(j+2)$ are entangled if
  we choose the coin state to be $|1\rangle$. With an increase in the number
  of steps, the state of the particle spreads in position space and the
  projection over one of the coin state reduces that state to a pure state, for which one may compute 
  spatial entanglement, according to the above prescription. Therefore, 
  with an increase in the number of steps, the spatial entanglement from a single-particle QW decreases.   

\subsubsection{Using many-particle quantum walk}
\label{mbqw}

We will extend the  study of evolution of spatial entanglement as the QW progresses on a many-particle system.   
\par
Let us first consider the analysis of first two steps of the Hadamard
walk ($\theta = \pi/4$ in Eq. (\ref{coin1})) on a three-particle
system with the initial state: 
\begin{equation}
 |\Psi_{ins}^{3p}\rangle = \bigotimes_{j=-1}^{+1}
  \left( \frac{|0\rangle + i|1\rangle}{\sqrt{2}} \right) \otimes
  |\psi_{j}\rangle. 
  \label{initialMBQWstate3p}
\end{equation}
We will label the three particles at positions $-1$, $0$ and
$1$ as ${\rm A}$, ${\rm B}$ and ${\rm C}$, respectively. Since evolution of these particles is 
independent, we write down the state after the first step as a tensor
product of each of the three particles: 
\begin{align}
 |\Psi^{3p}_{1}\rangle = W_{0, \theta, 0}^{\otimes 3}|\Psi_{ins}^{3p}\rangle 
  = \left[ \gamma  |0\rangle |-2  \rangle + \delta |1\rangle| 0 \rangle  \right]_{\rm A}  \otimes \left[ \gamma  |0\rangle |-1  \rangle + \delta |1\rangle| +1 \rangle \right]_{\rm B}  
        \otimes \left[ \gamma  |0\rangle |0
        \rangle + \delta |1\rangle| +2 \rangle  \right]_{\rm C},
  \label{threeparticle1step} 
\end{align}
where $\gamma = (1+i)/2$ and $\delta = (1-i)/2$.  After  two steps the 
tensor product of each of the three particles is given by
\begin{align}
  |\Psi^{3p}_{2}\rangle &= \left[ \gamma \left( \frac{|0\rangle |-3
        \rangle+|1\rangle|-1 \rangle}{\sqrt{2}} \right) + \delta
    \left( \frac{|0\rangle |-1 \rangle - |1\rangle|+1
        \rangle}{\sqrt{2}} \right) \right]_{\rm A} \nonumber \\ 
  &\otimes  \left[ \gamma \left( \frac{|0\rangle |-2
        \rangle+|1\rangle|0 \rangle}{\sqrt{2}} \right) + \delta \left(
      \frac{|0\rangle |0 \rangle - |1\rangle|+2 \rangle}{\sqrt{2}}
    \right) \right]_{\rm B} \nonumber \\ 
  &\otimes \left[ \gamma \left( \frac{|0\rangle |-1
        \rangle+|1\rangle|+1 \rangle}{\sqrt{2}} \right) + \delta \left(
      \frac{|0\rangle |+1 \rangle - |1\rangle|+3 \rangle}{\sqrt{2}}
    \right) \right]_{\rm C}. 
  \label{threeparticlestate} 
\end{align}
 By projecting this state into one of the coin states (we choose state
 $|0\rangle \otimes |0\rangle \otimes |0\rangle$) 
we can obtain a state of the lattice for which spatial
entanglement may be computed. Then the
state of
the lattice after projection and normalization is 
 \begin{align}
    |\Psi_{lat}\rangle =&  \gamma^3 \,
    |{\rm A}\rangle_{-3}|{\rm B}\rangle_{-2}|{\rm C}\rangle_{-1}  \nonumber \\
  &  + \gamma^2 \delta \left( \,
      |{\rm A}\rangle_{-3}|{\rm B}\rangle_{-2}|{\rm C}\rangle_{1}
      + |{\rm A}\rangle_{-3}|{\rm B}\rangle_{0}|{\rm C}\rangle_{-1}  
     +|{\rm AC}\rangle_{-1}|{\rm B}\rangle_{-2} \right) \nonumber \\
    & + \gamma \delta^2  \left( \,
      |{\rm A}\rangle_{-3}|{\rm B}\rangle_{0}|{\rm C}\rangle_{1} +
      |{\rm AC}\rangle_{-1}|{\rm B}\rangle_{0} 
     + |{\rm A}\rangle_{-1}|{\rm B}\rangle_{-2}|{\rm C}\rangle_{1} \right) \nonumber \\
    &  + \delta^3 \, |{\rm A}\rangle_{-1}|{\rm B}\rangle_{0}|{\rm C}\rangle_{1},
    \label{mblatticestate}
  \end{align}
 where $A,B$ and $C$ represent the particle labels and the subscripts represent
 the position labels. In a similar manner we can obtain
 $|\Psi_{lat}\rangle$ for a system with a large number of
 particles. Then the next task is to calculate the spatial
 entanglement. 
 
\section{Calculating spatial entanglement in a multipartite system} 
\label{mpent}

In a system with two particles, the state is separable if we can write
it as a tensor product of individual particle states, and
entangled if not. For a system with $M > 2$ particles,  a state is
said to be fully separable  if it can be written as 
\begin{equation}
  | \psi \rangle = | \phi_1 \rangle \otimes | \phi_2 \rangle \otimes
  \cdots | \phi_k \rangle, 
  \label{partialentangled}
\end{equation}
when $k=M$. $|\phi_i\rangle$ will then denote the state of the $i^{th}$
particle. When $k < M$ a state is said to be \emph{partially} entangled and
when $k=1$ the state will be fully entangled.  
\par
Rather than using the von Neumann entropy to quantify multipartite entanglement
of a given state $\rho$, one sometimes often prefers to consider purity, 
which corresponds (up to a constant) to linear entropy, that is the
first-order term in the expansion of the von Neumann entropy around its maxima, 
given by 
\begin{equation}
  E = \frac{d}{d-1} \left[ 1 - {\rm Tr} \rho^2 \right]
  \label{linentropy}
\end{equation}
for a $d$-dimensional particle Hilbert space \cite{PWK04}.
To quantify the entanglement of multipartite pure states, one measure
commonly, used is the  Meyer- Wallach (M-W) measure  \cite{MW02}.  It
is the entanglement measure of a single particle to the rest of the
system, averaged over the whole of the system and is given by 
\begin{equation}
  E_{MW} = \frac{d}{d-1} \left[ 1 - \frac{1}{L} \sum_{i=1}^{L} {\rm Tr} \rho_i^2 \right]
  \label{W-M}
\end{equation}
where $L$ is the system size and $\rho_i$ is the reduced density
matrix of the $i^{th}$ subsystem. The M-W measure does not  diverge with
increasing system size and is relatively easy to calculate. 
\par
In a multipartite QW the dimension at each lattice point,
after projection over 
one particular state of coin, is $2^M$ where $M$ is the number of
particles. Hence, the expression for entanglement will be
\begin{align}
E_{MW}(|\psi_{lat}\rangle) & =
\frac{2^M}{2^M-1}\left(1-\frac{1}{2t+M+1}\sum_{j=-(t+\frac{M}{2})}^{t+\frac{M}{2}}{\rm
    tr}\rho_j^2\right) 
    \label{mw}
\end{align}
where $t$ is the number of steps and $\rho_j$ is the reduced
density matrix of $j^{th}$ lattice point. The reduced density matrix $\rho_{j}$ can be written as
\begin{align}
\rho_j &= \sum_{k} p^j_k|k\rangle\langle k|
\end{align}
 where $|k\rangle$ is one of the  $2^M$ possible states available for
 a lattice point and $p^j_k$ can be calculated once we have the
 probability distribution of an individual particle on the lattice.
\par
Since we have $M$ distinguishable particles, we have $2^M$
configurations depending upon whether a given particle is present in
the lattice point or not after freezing the state of the
particle. This set of configurations forms the 
basis for a single-lattice point Hilbert space. Now we can calculate $p^j_k$, the probability of
$k^{th}$ configuration of a particle in the $j^{th}$ lattice point as follows.
Let us say $a_j^{(l_i)}$ is the probability of the $i^{th}$ particle to be
or not to be in the $j^{th}$ lattice point depending on
  $l_i$. If $l_i$ is $1$, 
then it gives us the probability of the particle to be in the lattice
point. If $l_i$ is $0$, then $a^{(l_i)}_j$ is the probability of a particle not to be in the lattice point, that is, $a_j^{(0)} =
  1-a_j^{(1)}$. Hence, we can write  
\begin{align}
p^j_k &= \prod_{i}a^{l_i}_j.
\end{align}
Once we have the probability of each particle at a given lattice position,  
the spatial entanglement can be conveniently calculated. 
Since the QW is a controlled evolution, one can obtain a
probability distribution of each particle over all lattice
positions. In fact, one can easily control the probability distribution 
by varying quantum coin parameters during the QW process
and hence the entanglement.
\par
\begin{figure}[ht]
\begin{center}
\includegraphics[width=9.5cm]{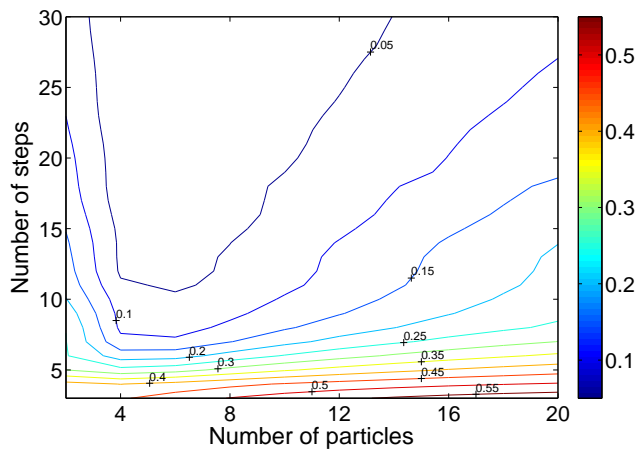}
\caption{(color online) Evolution of spatial entanglement with an increase in the number
  of steps of the QW for different number of particles in an
  open one-dimensional lattice chain.  The entanglement first
  increases and with further increase in the number of steps,
  the number of lattice positions exceeds the number of particles in
  the system resulting in the decrease of the spatial entanglement.  
The distribution is obtained by implementing the QW on particles
in the initial state $\frac{1}{\sqrt 2}(|0\ra + i |1\ra)$ and the Hadamard
operation $C_{0, \pi/4, 0}$ as quantum coin operation.} 
\label{eeqw}
\end{center}
\end{figure}
\begin{figure}[ht]
\begin{center}
\includegraphics[width=8.5cm]{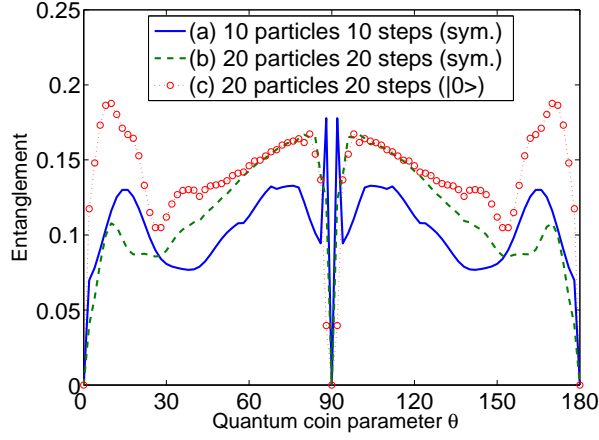}
\caption{(color online) Quantity of spatial entanglement for 10
  particles after 10 steps and 20 particles after 20 steps of the QW on a
  one-dimensional lattice using different 
  values of $\theta$ in the quantum coin operation $C_{0, \theta, 0}$.  
For (a) and (b), the distribution is for particles initially in the symmetric superposition state,
$\frac{1}{\sqrt{2}}(|0\rangle + i|1\rangle)$, and for (c) the particle's initial state is $|0\rangle$
 (will be the same for state $|1\rangle$). Quantity of entanglement is
 higher for $\theta$ closer to $0$ and $\pi/2$ compared to the
 intermediate value. We note that the asymmetric probability
 distribution due to an asymmetric initial state in case of (c)
 contributes for an increase in the quantity of spatial
 entanglement. When $\theta = \pi/2$, for every even number of steps
 of the QW, the system returns to the initial state where entanglement is
 $0$. Entanglement is 0 for $\theta= 0$.}  
\label{entqwtheta}
\end{center}
\end{figure}
\begin{figure}[ht]
\begin{center}
\includegraphics[width=8.5cm]{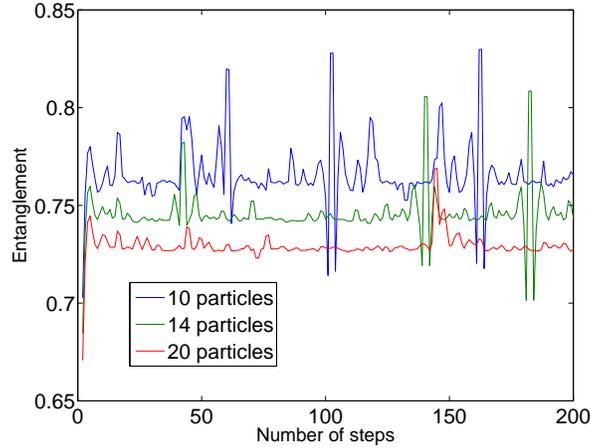}
\caption{(color online) Evolution of spatial entanglement for a system with
  different number of particles in a closed chain.  With an increase in
  the number of steps, the entanglement value remains close to  
asymptotic value with some peaks in between. The peaks can be
accounted for the crossover of leftward and rightward propagating
amplitudes of the internal state of the particle during the QW. 
The peaks are more for a chain with a smaller number of
particles. An increase in the number of particles in the system results
in the decrease of the entanglement value. The distribution is
obtained by using $\frac{1}{\sqrt{2}}(|0\ra + i|1\ra)$ as the initial
states of all particles and the Hadamard operation $C_{0, \pi/4,
  0}$ as quantum coin operation.} 
\label{Ering}
\end{center}
\end{figure}
\begin{figure}[ht]
\begin{center}
\includegraphics[width=8.5cm]{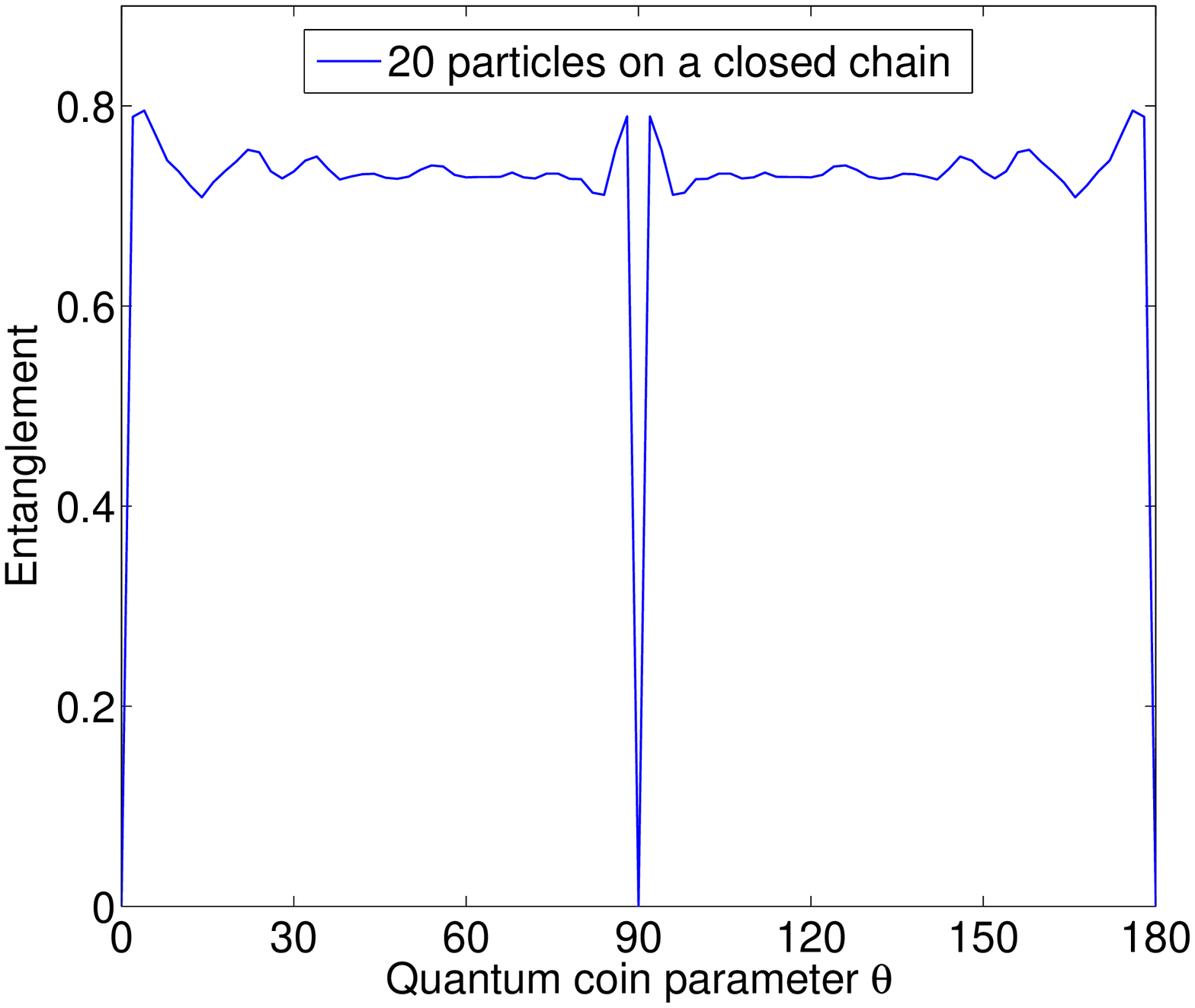}
\caption{(color online) Quantity of spatial entanglement for 20 particles on a closed
  chain after 20 steps of the QW using different values of
  $\theta$ in the quantum coin operation $C_{0, \theta, 0}$. The distribution is for 
  particles initially in the state $\frac{1}{\sqrt{2}}(|0\rangle + i|1\rangle)$. 
  Since the system is a closed chain, the QW does not expand the position Hilbert space, and
  therefore for all values of $\theta$ from $0$ to $\pi/2$ the
  entanglement value remains roughly uniform except for
  a small peak at smaller values of $\theta$.  For $\theta=0$
  when the number of steps equal to the number of particles, the amplitudes
  goes round the chain and returns to its initial state making the
  entanglement $0$ and for $\theta = \pi/2$, for every even number of
  steps of the QW, the system returns to the initial state where
  entanglement is again $0$.} 
\label{Eringtheta}
\end{center}
\end{figure}
\par
Fig. \ref{eeqw} shows the phase diagram of the spatial entanglement using a many-particle QW. 
Data for the phase diagram were obtained numerically by subjecting the many-particle system 
with different number of particles to the QW with increasing number of steps. 
The quantity of spatial entanglement was computed using Eq. (\ref{mw}). 
\par
Here, we have chosen the Hadamard operation $C_{0, \pi/4, 0}$ and
$\frac{1}{\sqrt 2}(|0\rangle  +  i |1\rangle)$ as the quantum coin
operation and initial state of the particles, respectively, for the evolution of the many-particle QW. 
To see the variation of entanglement for a fixed number of
particles with an increase in steps, we can pick a line parallel to the $y$
axis. That is, fix the number of particles and see the variation of entanglement
with the number of steps.
\par
In Fig. \ref{eeqw}, we see that for a fixed number of particles, the entanglement at first increases to some value before gradually falling. For $M=12$ we can note that the peak value is about $0.5$ before gradually falling.  With an increase in the number of steps of the QW, the 
number of lattice positions to which the particles evolve increases resulting 
in the decrease of the spatial entanglement (see Eq. (\ref{mw})). 
The decrease in entanglement before the number of steps is equal to the number of
particles should be noted. This is because for the Hadamard walk the
spread of a probability distribution after $t$ steps is between
$\frac{-t}{\sqrt 2}$  and $\frac{t}{\sqrt 2}$  \cite{CSL08}. 
\par
If we fix the number of steps and measure the entanglement by
increasing the number of particles in the system, the quantity of
spatial entanglement first decreases and then it starts increasing with
an increase in the number of particles. 
\par
To show the variation of spatial entanglement with the quantum coin
parameter $\theta$, we plot the spatial entanglement by varying the
parameter $\theta$ for a system with 10 particles after 10 steps of the QW 
and for a system with 20 particles after 20 steps of the QW in Fig.  \ref{entqwtheta}. 
In this figure, (a) and (b) are plots that use the symmetric
superposition state $\frac{1}{\sqrt 2}(|0\ra + i|1\ra)$ (unbiased QW)
as an initial state of all the particles, and (c) is the plot with all
the particles in one of the basis states $|0\ra$ or $|1\ra$ (biased QW)
as the initial state. We note that the quantity of entanglement is higher
for $\theta$ values closer to $0$ and $\pi/2$ and dips for values
in between for all the three cases.  
Biasing the QW, plot (c) shows
a slight increase in the quantity of entanglement compared to the
unbiased case, plot (b). A similar effect is seen by biasing the QW
using two parameters $\xi, \zeta$ in the coin operation $C_{\xi,
  \theta,\zeta}$ on particles initially in a symmetric superposition
state.   
\par
{\it Closed chain: } 
Since most physical systems considered for
implementation will be of a definite dimension, we extend our
calculations to one of the simplest examples of closed geometry,
an $n-$cycle. For a QW on an $n-$cycle, the shift operation,
Eq. (\ref{eq:condshift}), takes the form  
\begin{eqnarray}
\label{eq:condshift1}  S  =  |0\rangle  \langle 0|\otimes  \sum_{j  =
  0}^{n-1} |\psi_{j-1 \mbox{~mod~}n}\rangle  \langle \psi_{j} |
+|1\rangle  \langle 1 |\otimes \sum_{j =0}^{n-1} |\psi_{j+1 \mbox{~mod~}n}\rangle \langle \psi_{j}|. 
\end{eqnarray}
When we consider a many-particle system in a closed chain, with the number
of lattice positions equal to the number of particles $M$, the QW process
does not expand the position Hilbert space like it does on an open
chain (line). Therefore the spatial entanglement does not decrease at later times as it does for a walk on an open chain, but
remains close to the asymptotic value. Fig.  \ref{Ering} shows the evolution of entanglement for
a system with different number of particles in a closed chain. The
peaks seen in the plot can be accounted for by the crossover of the leftward and
rightward propagating amplitudes of the internal state of the particle
during the QW process. The frequency of the peaks is more for a smaller
number of particles (smaller closed chain).  Also, note that the
increase in the number of particles and the number of lattice points in
the closed cycle results in the decrease in spatial entanglement of
the system. 
\par
In Fig.  \ref{Eringtheta}, the value of spatial entanglement for 20
particles on a closed chain after 20 steps of the QW using different
values of $\theta$ in the quantum coin operation $C_{0, \theta, 0}$ is
presented.  For all values of $\theta$ from $0$ to $\pi/2$ the
entanglement value remains roughly uniform except for
the extreme values of $\theta$.  For $\theta=0$, the amplitude goes round the ring and
returns to its initial state making the spatial entanglement value $=
0$. For $\theta = \pi/2$, for every even number of steps of the QW,
the system returns to the initial state where spatial entanglement is
again $0$.  
\par
Therefore, spatial entanglement on a large lattice space can be
created, controlled and optimized for a maximum entanglement value by
varying the quantum coin parameters and number of particles in the
multi-particle QW.  

\section{Conclusion} 
\label{conc}

We have presented the evolution of spatial entanglement in a
many-particles system subjected to a QW process. By considering
many particle in the one-dimensional open and closed chain we have
shown that  spatial entanglement can be generated and controlled by
varying the quantum coin parameters, the initial state and the number of steps
in the dynamics of  the QW process. The spatial entanglement generated
can have a potential application in quantum information theory and
other physical processes.  
\bc
{\bf Acknowledgement} 
\ec

C.M.C is thankful to Mike and Ophelia Lezaridis for the financial support at IQC, 
ARO, QuantumWorks and CIFAR for travel support.  C.M.C also thank IMSc, Chennai, 
India for the hospitality during November - December 2008. S.K.G thanks Aiswarya Cyriac for the help in programming.


\end{document}